\documentclass[a4paper]{jpconf}
\usepackage{graphicx}
\usepackage{listings}
\usepackage{color}

\begin{document}
\title{General purpose lattice QCD code set
Bridge++ 2.0 for high performance computing}

\author{%
Yutaro Akahoshi$^1$, %
Sinya Aoki$^1$, %
Tatsumi Aoyama$^2$, %
Issaku Kanamori$^3$, %
Kazuyuki Kanaya$^4$, %
Hideo Matsufuru$^2$, %
Yusuke Namekawa$^5$, %
Hidekatsu Nemura$^1$, %
Yusuke Taniguchi$^4$}

\address{$^1$ Yukawa Institute for Theoretical Physics, Kyoto University, Japan}
\address{$^2$ High Energy Accelerator Research Organization (KEK), Japan}
\address{$^3$ RIKEN Center for Computational Science (R-CCS), Japan}
\address{$^4$ University of Tsukuba, Japan}
\address{$^5$ Department of Physics, Kyoto University, Japan}

\ead{kanamori-i@riken.jp}

\begin{picture}(0,0)(0,0)
 \put(400,300){KUNS-2899}
\end{picture}
\begin{abstract}

Bridge++ is a general-purpose code set for a numerical simulation of
lattice QCD aiming at a readable, extensible, and portable code
while keeping practically high performance.
The previous version of Bridge++ is implemented in double precision
with a fixed data layout.
To exploit the high arithmetic capability of new processor architecture,
we extend the Bridge++ code so that optimized code is available
as a new branch, \textit{i.e.}, an alternative to the original code.
This paper explains our strategy of implementation and displays
application examples to the following architectures and systems:
Intel AVX-512 on Xeon Phi Knights Landing, Arm A64FX-SVE on
Fujitsu A64FX (Fugaku), NEC SX-Aurora TSUBASA, and GPU cluster
with NVIDIA V100.

\end{abstract}

\section{Introduction}

Bridge++\footnote{\texttt{https://bridge.kek.jp/Lattice-code/}}
is a code set for numerical simulations of lattice
QCD\footnote{Basics of lattice QCD are covered by many text textbooks, \textit{e.g.}, 
\cite{Knechtli:2017sna}.}, 
designed on the object-oriented programming and described
in the C++ language.
A goal of the project is to develop a readable, extensible, and portable code
set with sufficiently high performance.
When the development was launched in 2009, our major target platforms
were parallel scalar systems represented by IBM Blue Gene/Q.
Recent supercomputers, however, adopt a variety of architecture:
multi-core parallel machines with wide SIMD (A64FX and Intel
processors), and clusters with accelerator devices such as GPUs,
PEZY-SC, and vector processors (NEC SX-Aurora).
Soon after the first public release of Bridge++ in 2012
\cite{Ueda:2014rya},
we had started to investigate possible extensions of our code to exploit these
new architectures while keeping the readability and portability,
as well as to develop tuning techniques for them
\cite{Motoki:2014a,Matsufuru:2015a,Aoyama:2016a,Kanamori:2017a,
Kanamori:2018hwh,iccsa21}.
Recently we have constructed a framework to incorporate
the tuned codes as an alternative part to the previously developed
Bridge++ code, and decided to release it as version 2.0.

In this paper, we describe the fundamental structure of this
updated code set with several examples of application to recent
architectures.
In the next section, after a brief introduction of the lattice QCD
and its bottleneck, we describe the structure of Bridge++ ver.2.0
code set.
In Sec.~\ref{sec:Implementation_and_performance} 
we show %several examples of application to recent architecture with 
the performance
measured on available systems: the SIMD architectures (Intel AVX-512 and
Arm A64FX), vector architecture (NEC SX-Aurora TSUBASA),
and GPU devices with NVIDIA V100.
The last section is our conclusion.
Some details of the implementation have been reported in Ref.~\cite{iccsa21}
together with that of a multi-grid solver.

\section{Functionality and Code Structure}

The lattice QCD is a field theory formulated on a four-dimensional
Euclidean lattice.
It consists of fermion (quark) fields and a gauge (gluon) field.
The latter mediates the strong interaction among quarks and are
represented by
`a link variable', $U_\mu(x)\in SU(3)$, where $x=(x_1,x_2,x_3,x_4)$
stands for a lattice site and $\mu =1$--$4$ is the spacetime direction.
In numerical simulations the lattice size is finite:
$x_\mu =1,2,\dots,L_\mu$.
The fermion field is represented as a complex vector on lattice sites,
which carries 3 components of `color' and 4 components of `spinor',
thus in total 12, degrees of freedom on each site.
We set the lattice spacing $a=1$ throughout this paper.

The standard steps of lattice QCD simulations are as follows.
One first generates gauge configurations $\{ U_\mu(x)\}$ with
a Monte Carlo method.
Then physical observables are measured on $\{ U_\mu(x)\}$, and
their averages give the expectation values.
Typical observables, such as hadronic matrix elements, are composed
of fermion propagators, which are solutions of the linear equation
for a fermion matrix $D[U]$.
Since such a linear equation must be solved repeatedly also during
the generation of $\{ U_\mu(x)\}$, it is one of the main bottlenecks of
lattice simulations.
The fermion matrix, explicitly exemplified below,
represents the interaction among quarks and gluons and is in general
a large sparse matrix of a rank proportional to the lattice volume.
The discretized fermion matrix has a variety since
the requirement on $D[U]$ is to coincide with that of QCD
only in the continuum limit (the lattice spacing $a\rightarrow 0$).
In addition, efficient solver algorithms also have a variety
depending on the system size, the condition number of the matrix,
and so on.

Our code set Bridge++ is intended to cover a wide range of
measurements, fermion matrices, and algorithms.
It is parallelized with MPI and the time consuming parts are
multi-threaded with OpenMP.
The data structure initially adopted was in a fixed style with double precision.
For further optimization to up-to-date architectures, we decided to replace the time consuming
part with an alternative code leaving the other parts unchanged.
The original code, called `core library',
provides building tools,
reference results, and measurements that require less performance.
On the other hand, the `alternative code' offers the same
functionality as that of the core library with higher flexibility to achieve better performance.
It is possible to use multiple branches of the alternative code
simultaneously if required.

The `alternative code' is composed of the classes in two categories.
One is those directly manipulate the data and thus are
performance-sensitive, such as the fermion matrices.
These classes are implemented and optimized in a way specific to
each architecture.
The other category contains the algorithms generically described
by the C++ template.
This structure allows incremental adoption to a new architecture.
One only needs to implement the required operators and instantiate
the desired algorithms.

\begin{figure}
 {
\definecolor{orange}{cmyk}{0, .55, .95, 0.1} 
\small
 \lstset{language=C++, %
 morekeywords={start, wait},%
 basicstyle=\ttfamily\fontsize{7pt}{9pt}\selectfont,%
 commentstyle=\textit,%
 stringstyle={\ttfamily \color[cmyk]{0,1,0,0}},%
 columns=fullflexible,%
 frame=tb,%
 xleftmargin=2em,%
 framexleftmargin=2em,%
 showstringspaces=false,%
 numbers=left, stepnumber=1,numberstyle=\tiny,%
 numbersep=1em,%
 moredelim=[is][\color{blue}]{[\#alt\#}{\#]},
 moredelim=[is][\color{orange}]{[\#core\#}{\#]},
 moredelim=[is][\color{red}]{[\#template\#}{\#]},
 }
 \begin{lstlisting}
  // prepare a mixed precision solver for architecture SIMD
  //    details of the fermion op.: params_fopr;  details of the solver: params_solver
  [#alt#Fprop_Standard_lex_alt_Mixedprec#]<[#template#AField<double,SIMD>, AField<float,SIMD>#]> [#alt# fprop(#][#core# params_fopr, params_solver #][#alt#)#];
  [#alt#fprop.set_config(#][#core#U.get()#][#alt#)#];  // set the gauge field
  [#alt#fprop.set_mode(#]"D"[#alt#)#];                 // mode is D, Ddag, DdagD,...

  // source and solution are from core library 
  [#core#Source source(#]"Local"[#core#);  source.set_parameters(params_source)#];
  [#core#std::vector<Field_F> sq(#]Nc * Nd[#core#)#];  // Nc=3, Nd=4 : propagator (= 12 solution vectors)
  [#core#Field_F b#];

  int    nconv;  double diff;
  for (int id = 0; id < Nd; ++id) {
    for (int ic = 0; ic < Nc; ++ic) {
      int idx = ic + Nc * id;
      [#core#sq[idx].set(#]0.0[#core#)#];
      [#core#source.set(b, #]idx[#core#)#]; // set the source
      [#alt#fprop.invert(#][#core#sq[idx]#], [#core#b#],   nconv, diff[#alt#)#];  // solve the linear equation
    }
 }
 \end{lstlisting}
}
\caption{
A sample code to calculate a fermion propagator with a mixed precision
solver.  The blue-colored codes are from the `alternative' code with the red-colored template parameter, while the orange-colored objects are from the `core library'.}
\label{fig:code_example}
\end{figure}

Figure~\ref{fig:code_example} shows a sample code which employs
the extended Bridge++ with a branch for SIMD architecture.
In line-3, {\tt AField<double,SIMD>} indicates a class of
field object in double precision for the `SIMD' branch.
The template class {\tt Fprop\_Standard\_lex\_alt\_Mixedprec}
determines a fermion propagator by a mixed precision solver
algorithm.
The sets of parameters for the fermion matrix and the linear equation
solver are stored in the objects {\tt params\_fopr} and
{\tt params\_solver},
respectively, including the types of fermion and algorithm.
Once the object `{\tt fprop}' is instantiated, the
solution of the linear equation is determined and returned in
the format of the core library, {\tt Field\_F}.
One does not need to modify observables that use the solution `{\tt sq[idx]}'
to use the `alternative' codes.

\section{Implementation and Performance for each Architecture}
\label{sec:Implementation_and_performance}

We describe our code implementation and performance
results for several architectures.
While the code is available in both double precision and single
precision, we mainly describe the latter since it plays a main
role in multi-precision solver algorithms.
We quote weak scaling behavior
of the performance of matrix-vector multiplication without details
of the tuning and execution setup.

A fermion matrix $D[U]$ acts on a fermion vector $\psi(x)$.
As a typical example, we examine the $O(a)$-improved
Wilson fermion matrix, also called clover fermion matrix,
\begin{equation}
D_{x,y} = [1 + F(x)] \delta_{x,y} - \kappa \sum_{\mu=1}^4 
  \left[ (1-\gamma_\mu) U_\mu(x) \delta_{x+\hat{\mu}, y}
       + (1+\gamma_\mu) U^\dag_\mu (x-\hat{\mu}) \delta_{x-\hat{\mu},y} \right] ,
\label{eq:fermion_matrix}
\end{equation}
where $x$, $y$ are lattice sites, $\hat{\mu}$ the unit vector
along $\mu$-th axis, and the hopping parameter
$\kappa = 1/(8+2m_0)$ related to the quark mass $m_0$.
The link variable $U_\mu(x)$ is a $3\times 3$ complex matrix acting
on the color and $\gamma_\mu$ is a $4 \times 4$ matrix acting on
the spinor degrees of freedom.
$F(x)$ is a $12\times 12$ Hermitian matrix made of the link variables,
and helps to reduce the finite lattice spacing artifact.
Thus $D$ is a complex matrix of the rank $4\cdot 3  L_x L_y L_z L_t$.
The boundary condition, such as a periodic boundary, is imposed in each
direction.

\subsection{SIMD architectures: Intel AVX-512 and Fujitsu A64FX}

We start with the implementation for two SIMD architectures:
Intel AVX-512 and Armv8.2-A with SVE (Scalable Vector Extension).
Intel AVX-512 is the latest SIMD extension of x86 instruction
set architecture with the 512-bit SIMD length.
It is available on recent Xeon processors as well as Xeon Phi
Knights Landing.
The Armv8.2-A with SVE is adopted by the Fujitsu A64FX processor
for the Fugaku supercomputer.
While SVE enables variable SIMD length, currently 512-bit length
is available on A64FX.

\begin{figure}
\centering
\includegraphics[width=0.37\linewidth]{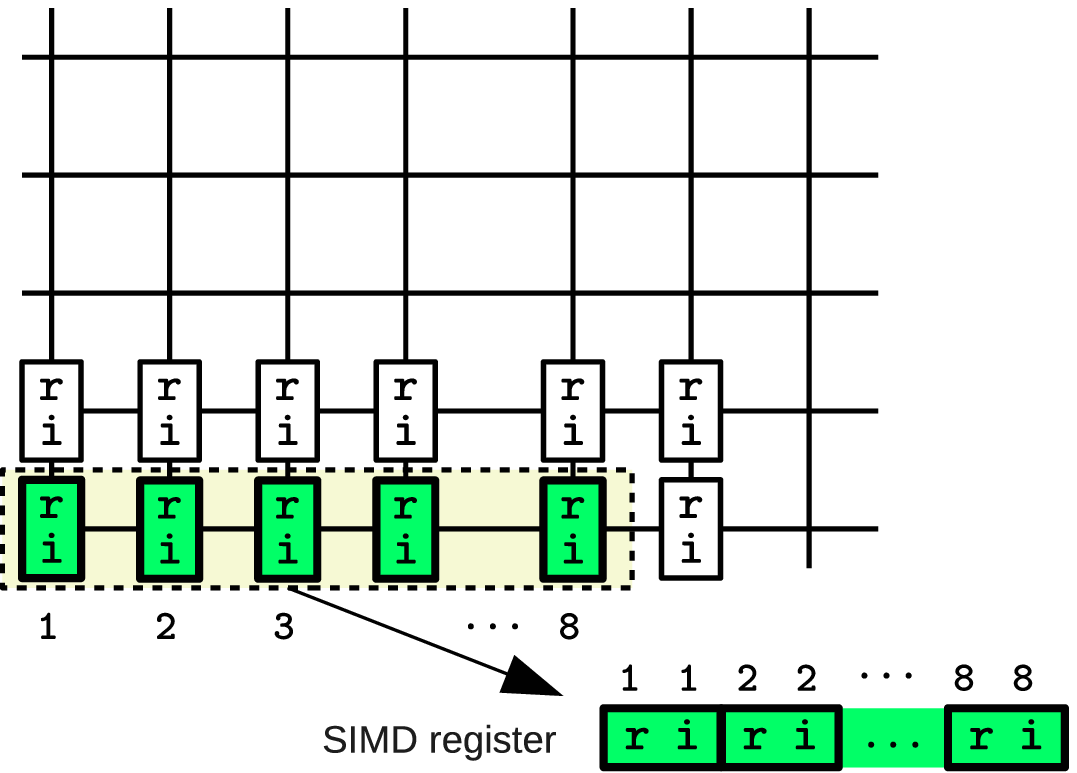}
\hspace{0.5cm}
\includegraphics[width=0.39\linewidth]{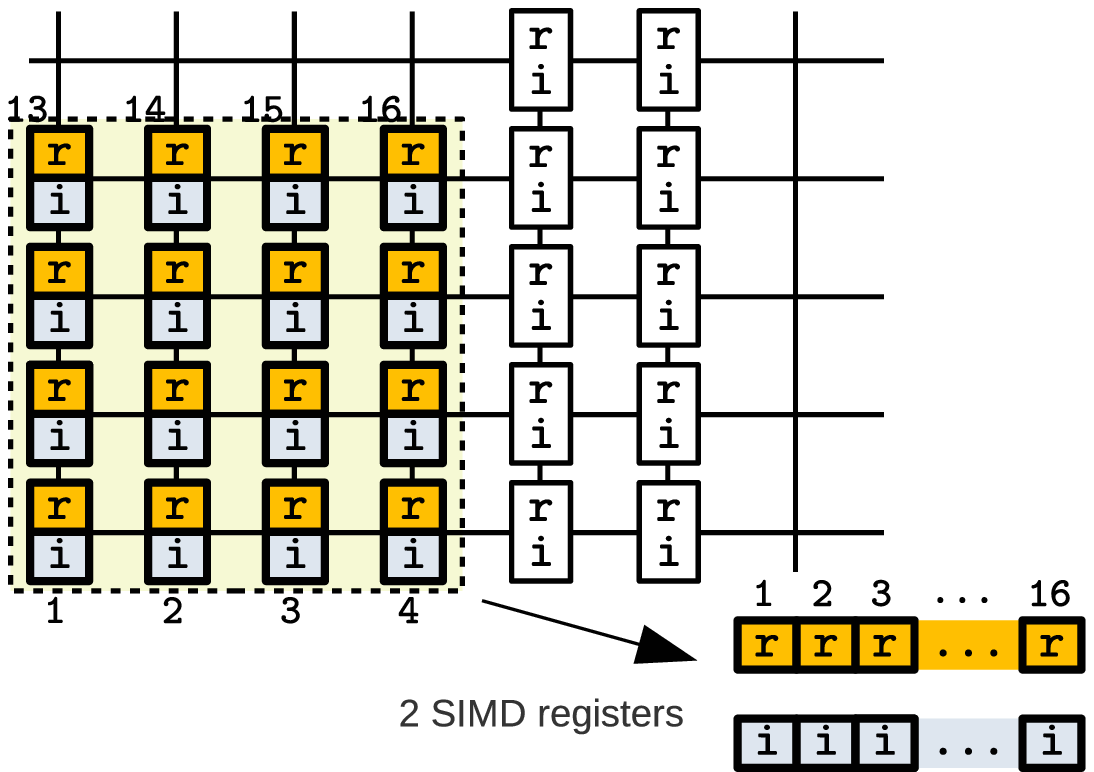}
\caption{SIMD data layout we adopted for AVX-512
(left) and A64FX (right) architectures.
}
\label{fig:SIMD_layout}
\end{figure}

Although in both architectures the SIMD length corresponds
to 16 single precision floating numbers, 
the efficient way of packing variables into
the SIMD unit depends on the structure of arithmetic
operation units.
In the case of AVX-512, we pack 8 complex numbers which are
consecutive in the $x$-direction as displayed in the left panel of
Fig.~\ref{fig:SIMD_layout}.
This requires that the lattice size in $x$-direction must be a
multiple of 8.
The details of the tuning with the AVX-512 instruction set were presented
in \cite{Kanamori:2018hwh}.
For Armv8.2-A-SVE, we adopt a different packing: as depicted in the right
panel of Fig.~\ref{fig:SIMD_layout}, real and imaginary parts
are separately stored.
This packing shows better performance, found through the development
of QCD Wide SIMD (QWS) library \cite{Ishikawa:2021iqw} as a product of the Post-K
co-design project.
With this layout, the field variables on 16 sites are executed in parallel.
To keep flexibility in choosing the lattice size, we pack variables in the $x$-$y$ plane into a SIMD
vector, while a one-dimensional packing was adopted in QWS.
Therefore we prepare two branches of code for the AVX-512 and
A64FX architectures.
Since the branch for A64FX also intends to call the QWS library,
we need to convert the data layout and the physical convention that
are different in Bridge++ and QWS.

\begin{figure}
 \centering
\includegraphics[width=0.45\linewidth]{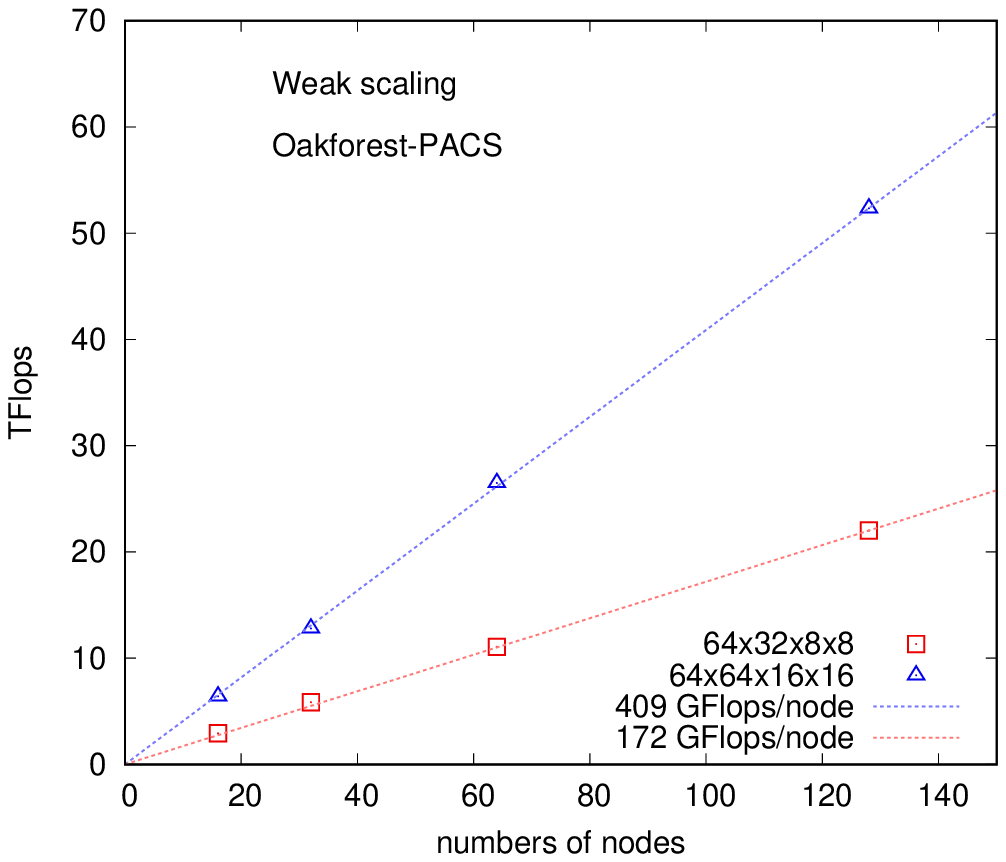}%
\includegraphics[width=0.45\linewidth]{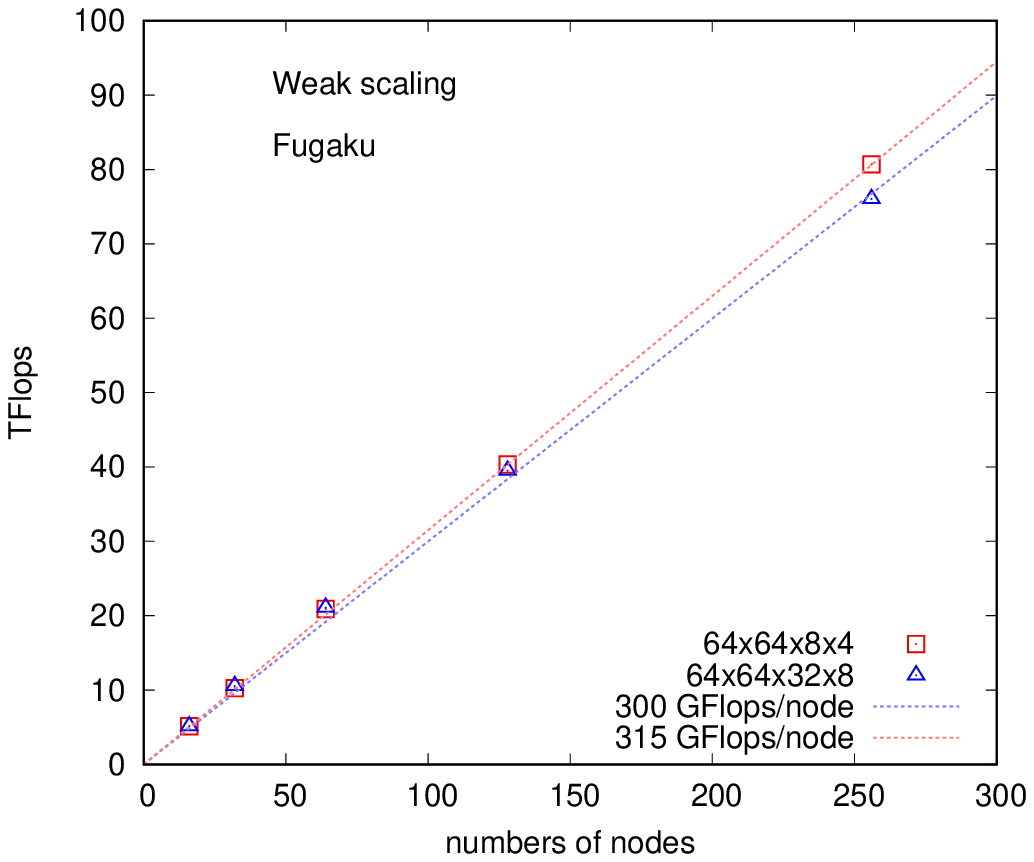}%
\caption{Weak scaling of the performance of 
the clover matrix multiplication on Oakforest-PACS (left) and
Fugaku (right). The dotted lines are scaling from the largest numbers of node.}
\label{fig:weak_scaling_OFP_Fugaku}
\end{figure}

We measure the performance of our code on the Oakforest-PACS system
at JCAHPC, a cluster composed of the Intel Xeon Phi Knights Landing
processors, and the supercomputer Fugaku at RIKEN.
In Fig.~\ref{fig:weak_scaling_OFP_Fugaku}, weak scaling behavior of
the Clover matrix multiplication is displayed for these SIMD
architectures.
The results show good scaling on both systems, although Oakforest-PACS
shows significant dependence on the lattice size per node 
as the smaller local volume is too small to hide the neighboring communication overlapped with the computation in bulk.
The peak performance of each node is about 6 TFlops on both
the systems so that the current sustained performance is around 5 \%.
This is reasonable performance for practical simulations, though
there is still room for improvement.

\subsection{Vector architecture: NEC SX-Aurora TSUBASA}

The latest vector architecture of NEC is adopted in the NEC
SX-Aurora TSUBASA system whose vector length is 256 in units of double precision floating-point number.
The best performance is obtained for the loop size with multiples
of this vector length, and thus we assign this vector index
to the lattice sites.
The data layout suitable to this setup is so-called
Structure of Array (SoA) layout.
We rearrange the data so that the site degrees of freedom
are consecutively stored on the memory, including the padding to
avoid the bank conflict.

We measure the performance on the NEC SX-Aurora TSUBASA at KEK.
Each node of the system is composed of one Vector Host (Intel Xeon
processor) and eight Vector Engines (VE).
Each VE has 2.42 TFlops of the peak performance and 1.2 TB/s memory
bandwidth, indicating the byte-per-flop of 0.5 as an advantage of
this architecture.
One can execute MPI jobs in units of core in VE (8 cores/VE).

\begin{figure}
 \centering
\includegraphics[width=0.45\linewidth]{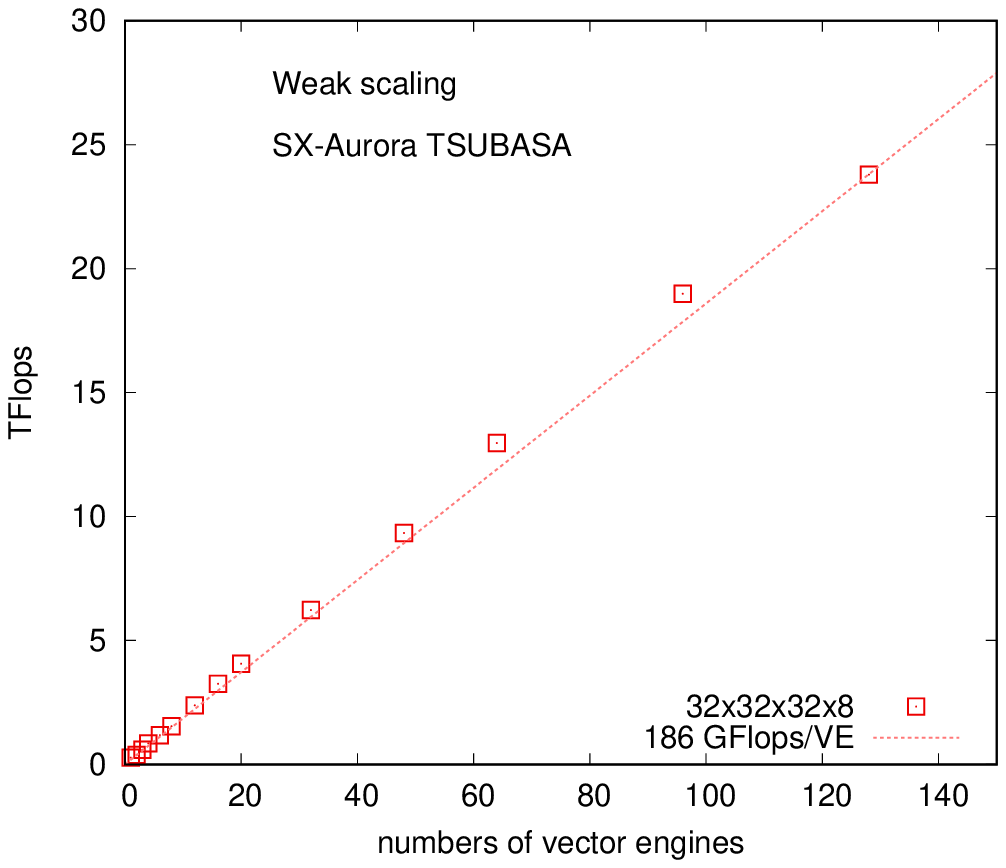}
\includegraphics[width=0.45\linewidth]{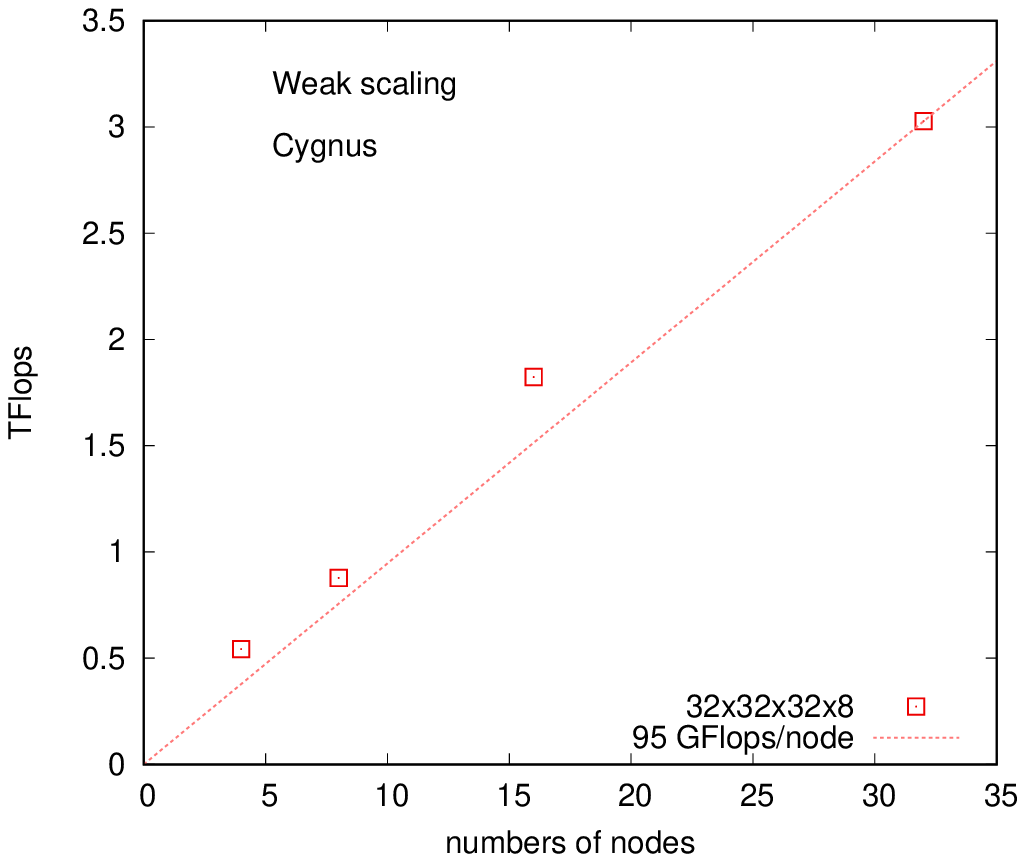}%
\caption{
Weak scaling behavior of the performance of matrix-vector
multiplication for the Wilson matrix on SX-Aurora TSUBASA (left)
and the Clover matrix on the Cygnus system (right).
The dotted lines are scaling from the largest numbers of nodes or vector engines.
}
\label{figs:weak_scaling_SX-Aurora_Cygnus}
\end{figure}

In the left panel of Fig~\ref{figs:weak_scaling_SX-Aurora_Cygnus},
we show the weak scaling behavior of the Wilson matrix multiplication
(setting $F(x)=0$) since the clover fermion matrix is in preparation.
While sufficient performance is obtained on the single core,
the performance decreases on multi-process for unspecified reasons.
Results on multiple VE show good scaling behavior.
The vector instruction ratio is 99.90\% with average vector length 256.0.

\subsection{GPU Cluster}

On the clusters with accelerator devices, one needs to offload
the data and tasks to the devices employing an offloading scheme.
We implement such a code using OpenACC, which is a widely used
directive-based framework.
By calling an interface object, the data layout is rearranged to
that respects so-called coalesced access, and the data are
automatically transferred between the host and device.
Thus one can generally compose the algorithms using these classes
without worrying about the data transfer.
The tasks to be offloaded are specified by OpenACC directives.
We extract such kernel functions and collect them as a library
so that one can replace them with more optimized functions
with CUDA or OpenCL if necessary.

As an example of GPU cluster, we use the Cygnus system at the University
of Tsukuba.
Each node of Cygnus is composed of two Intel Xeon processors
and four NVIDIA Tesla V100 GPUs.
Each V100 GPU has 5120 CUDA cores which amount to 14 TFlops for
FP32 arithmetics.
While a GPU has high arithmetic performance, bottlenecks are
the data transfer between the host and device (PCIe 3.0 with 16 lanes)
and between the device global memory and device cores
(900 GB/s for V100).
Together with the Infiniband connection among nodes, one needs
to choose the parameters considering the heterogeneous structure
of the GPU cluster.

In the right panel of Fig.~\ref{figs:weak_scaling_SX-Aurora_Cygnus},
we show the weak scaling of clover fermion matrix multiplication
on the Cygnus system.
The sustained performance is governed by the device memory bandwidth.
Increasing the number of nodes, the overhead of inter-node
communication becomes sizable.

\section{Conclusions}

We presented features of our forthcoming major update of Bridge++ that incorporates
an optimized code for recent architectures as an alternative
to the original code.
We described our fundamental strategy of implementation
and displayed several examples in practical application.
These results demonstrate that our framework indeed works with
sufficient performance for practical application.
While we only showed sustained performance for the Wilson and
Clover fermion matrices, other types of fermion and a variety of
algorithms including multi-grid solvers and eigenvalue
solvers are available.
Now we are in the final stage toward the public release of
the Bridge++ version 2.0.

\ack

The authors would like to thank the members in the lattice QCD working
group in FS2020 project for post-K computer.
Numerical simulations were performed on the Oakforest-PACS and Cygnus
systems through Multidisciplinary Cooperative Research Program in CCS,
University of Tsukuba (16a43, 17a41, ID14(FY2018), ID43(FY2019), ID79(FY2020)),
supercomputer Fugaku at RIKEN Center for Computational Science,
and SX-Aurora TSUBASA system at KEK through Particle, Nuclear, and
Astro Physics Simulation Program (2019-T003, 2020-002, 2021-006).
Some parts of the code development were performed on the supercomputer
`Flow' at Information Technology Center, Nagoya University, and
Yukawa Institute Computer Facility.
This work is supported by JSPS KAKENHI (%
JP20K03961,
JP21K03553),
the MEXT as `Program for Promoting Researches on the Supercomputer
Fugaku' (Simulation for basic science: from fundamental laws of
particles to creation of nuclei) and
`Priority Issue 9 to be Tackled by Using the Post-K Computer' (Elucidation of the Fundamental Laws and Evolution of the Universe),
and Joint Institute for Computational Fundamental Science (JICFuS).

\section*{References}

\bibliography{ccp2021}

\providecommand{\newblock}{}
\begin{thebibliography}{1}
\expandafter\ifx\csname url\endcsname\relax
  \def\url#1{{\tt #1}}\fi
\expandafter\ifx\csname urlprefix\endcsname\relax\def\urlprefix{URL }\fi
\providecommand{\eprint}[2][]{\url{#2}}

\bibitem{Knechtli:2017sna}
Knechtli F, G\"unther M and Peardon M 2017 {\em {Lattice Quantum
  Chromodynamics: Practical Essentials}\/} SpringerBriefs in Physics (Springer)
  ISBN 978-94-024-0997-0, 978-94-024-0999-4

\bibitem{Ueda:2014rya}
Ueda S, Aoki S, Aoyama T, Kanaya K, Matsufuru H, Motoki S, Namekawa Y, Nemura
  H, Taniguchi Y and Ukita N 2014 {\em J. Phys. Conf. Ser.\/} {\bf 523} 012046

\bibitem{Motoki:2014a}
Motoki S {\em et~al.\/} 2014 {\em Procedia Computer Science\/} {\bf 29}
  1701--1710

\bibitem{Matsufuru:2015a}
Matsufuru H {\em et~al.\/} (Bridge++ Project) 2015 {\em Procedia Computer
  Science\/} {\bf 51} 1313--1322

\bibitem{Aoyama:2016a}
Aoyama T, Ishikawa K~I, Kimura Y, Matsufuru H, Sato A, Suzuki T and Torii S
  2016 {\em Procedia Computer Science\/} {\bf 80} 1418--1427

\bibitem{Kanamori:2017a}
Kanamori I and Matsufuru H 2017 {\em Proceedings of the Fifth International
  Symposium on Computing and Networking\/} {\bf CANDAR'17} 66

\bibitem{Kanamori:2018hwh}
Kanamori I and Matsufuru H 2018 {\em Lecture Note on Computer Science\/} {\bf
  10962} 456--471

\bibitem{iccsa21}
Kanamori I, Ishikawa K~I and Matsufuru H 2021 {\em Lecture Note on Computer
  Science\/} {\bf 12953} 218--233

\bibitem{Ishikawa:2021iqw}
Ishikawa K~I, Kanamori I, Matsufuru H, Miyoshi I, Mukai Y, Nakamura Y, Nitadori
  K and Tsuji M 2021  (\textit{Preprint} \eprint{2109.10687})

\end{thebibliography}

\end{document}